\documentclass[epj]{svjour}
\usepackage{amsmath}
\usepackage{amsfonts}
\usepackage{amssymb}
\usepackage{graphics}
\usepackage{epsfig}

\begin{document}
\title{Integrable cosmological models with an additional scalar field}
\author{Vsevolod~R.~Ivanov\inst{1} \and Sergey~Yu.~Vernov\inst{2}
}                     % Do not remove
\titlerunning{Integrable cosmological models with an additional scalar field}
\authorrunning{V.R.~Ivanov, S.Yu.~Vernov}

\institute{Physics Department, Lomonosov Moscow State University, Leninskie Gory 1, 119991, Moscow, Russia, \email{vsvd.ivanov@gmail.com} \and
Skobeltsyn Institute of Nuclear Physics, Lomonosov Moscow State University, Leninskie Gory~1, 119991, Moscow, Russia, \email{svernov@theory.sinp.msu.ru}}

\date{Received:  / Revised version: date}
\abstract{We consider modified gravity cosmological models that can be transformed into two-field chiral cosmological models by the conformal metric transformation.
For the $R^2$ gravity model with an additional scalar field and the corresponding two-field model with the cosmological constant and nonstandard kinetic part of the action, the general solutions have been obtained in the spatially flat FLRW metric. We analyze the correspondence of the cosmic time solutions obtained and different possible evolutions of the Hubble parameters in the Einstein and Jordan frames.
 % end of PACS codes
} %end of abstract

%\keywords{Chiral cosmological model, modified gravity, exact solutions}

\maketitle

\section{Introduction}

Modified gravity models are actively investigated~\cite{Nojiri:2010wj,Clifton:2011jh,Capozziello:2011et,Nojiri:2017ncd,CANTATA:2021ktz}.
An important class of these models that includes $F(R)$ gravity models and models with a nonminimal coupling between the scalar field and the curvature is characterized by the possibility to construct the mathematically equivalent General Relativity models with a scalar field by the conformal metric transformation.
In other words, these models can be considered in the Einstein frame as models with one scalar field.

At present, a lot of modified gravity cosmological models with an additional scalar field have been proposed~\cite{Gottlober:1993hp,Qiu:2014apa,Kaneda:2015jma,delaCruz-Dombriz:2016bjj,Karam:2018mft,Canko:2019mud,Odintsov:2019evb,Mitsopoulos:2021qpu,Oikonomou:2021msx,Wang:2017fuy,Ema1,He:2018gyf,Ema2,Gorbunov:2018llf,Gorbunov:2019,He:2020ivk,Ema3,Cheong:2019vzl,Gundhi:2020kzm,Gundhi:2020zvb,SBB1989,Ferrara:2010in,Kaiser:2012ak,Greenwood:2012aj,Dubinin:2017irg,Dubinin:2017irqA}. In particular, inflationary models of the $R^2$ gravity with the Higgs-like boson~\cite{Wang:2017fuy,Ema1,He:2018gyf,Ema2,Gorbunov:2018llf,Gorbunov:2019,He:2020ivk,Ema3,Cheong:2019vzl,Gundhi:2020kzm,Gundhi:2020zvb} as well as inflationary models with multiple scalar fields nonminimally coupled to the curvature term~\cite{SBB1989,Ferrara:2010in,Kaiser:2012ak,Greenwood:2012aj,Dubinin:2017irg,Dubinin:2017irqA} are actively studied. Note that the $R^2$ term arises as a quantum correction when inflationary models with scalar fields are considered~\cite{Steinwachs:2011zs,Salvio:2015kka,Elizalde:2015nya,Salvio:2021lka}. One of the motivations to consider
the $R^2$ gravity with the Higgs-like boson is the possible production of primordial black holes in such types of models~\cite{Cheong:2019vzl,Gundhi:2020kzm,Gundhi:2020zvb}. Also, we want to mention the Higgs-dilaton cosmological model that has been proposed to describe simultaneously an inflationary expansion in the early Universe and a dark energy dominated stage responsible for the present-day acceleration~\cite{Bezrukov:2012hx}.

Unlike models with a single scalar field nonminimally coupled to gravity as well as $F(R)$ gravity models, models with multiple scalar fields nonminimally coupled with gravity cannot be transformed to models with minimally coupled scalar fields with the canonical kinetic terms in the generic case~\cite{Kaiser:2010ps}. After the metric transformation, one obtains models with a nonstandard kinetic part, so-called chiral cosmological models~\cite{Chervon:1995jx,DiMarco:2002eb,Chervon:2014dya,Braglia:2020fms,Braglia:2020eai,Anguelova:2020nzl}. Mathematical properties of such models
attract a lot of attention~\cite{Chervon:1995jx,Chervon:2014dya,Paliathanasis:2018vru,Christodoulidis:2019mkj,Paliathanasis,Chervon:2019nwq,Anguelova:2019omg,Giacomini:2020grc,Zhuravlev:2020ugb,Socorro:2020nsm,Fomin:2020woj,Paliathanasis:2020sfe,Paliathanasis:2021fxi}.

In spite of the great success of numerical and approximation methods, exact analytical solutions play an important role in studying evolution equations and in the investigation of some important qualitative features of cosmological models with scalar fields~\cite{Chimento:1998ju,Arefeva:2004odl,Andrianov:2006hx,Fre,Chervon:2017kgn,Chervon:2019sey,Faraoni:2021opj}. Most of the results of the exact integration of cosmological models with scalar fields are connected with one-field cosmological models~\cite{Fre,Maciejewski:2008hj,Kamenshchik:2013dga,Boisseau:2015hqa,Pol1,Kamenshchik:2015cla}. A method for constructing integrable models with nonminimally coupled scalar fields by using the interrelation between the Jordan and Einstein frames has been proposed for one-field models in Ref.~\cite{Kamenshchik:2013dga}. Sometimes, the integrability of cosmological models with nonminimal coupling is more apparent, than the integrability of the corresponding models in the Einstein frame~\cite{Boisseau:2015hqa,Pol1,Kamenshchik:2015cla}.

The generalization of this analysis on the case of a few scalar fields is not straightforward, because the conformal transformation of the metric changes the form of kinetic terms of the scalar fields.
So, to find integrable modified gravity models with an additional scalar field it can be useful to study the corresponding chiral cosmological model.
The integrability of many cosmological models has been proved by solving evolution equations with a suitable parametric time~\cite{Fre}, but it is not clear how to find this parametric time. In our paper, we show that for some modified gravity models the suitable parametric time is the cosmic time in the Einstein frame. We propose the method in which the parametric time is one and the same in both frames.

In this paper, we study chiral cosmological models with two scalar fields and the cosmological constant. The absence of the scalar field potential allows us to get the behaviour of the Hubble parameter in the analytic form, namely, in the form of hyperbolic tangent or cotangent. We also get scalar fields in the analytic form. The knowledge of this general solution allows us to get the general solution of the initial modified gravity models with an additional standard or phantom scalar field in the parametric time. After this, using the time transformation, we get solutions in the cosmic time. It is important to check the existence of other solutions to the initial modified gravity model.

The most famous $F(R)$ gravity model, the Starobinsky $R^2$ inflationary model~\cite{Starobinsky80,Starobinsky82,Mijic:1986iv,Maeda88,Ketov:2010qz,Bezrukov:2011gp},  includes both $R^2$ term and the standard Hilbert-Einstein term. Adding to this model the cosmological constant, one can obtain the model with exact cosmological solutions~\cite{Vernov:2019ubo}. The integrability of the Starobinsky model as well as the integrability of $R+R^n$ and pure $R^n$ cosmological models are actively investigated by the singularity analysis~\cite{Paliathanasis:2016tch} and other methods~\cite{Carloni:2004kp,Faraoni:2017ecj,Banijamali:2019gry}. It is known that a pure $R^2$ model is integrable~\cite{Paliathanasis:2016tch}. At the same time, the integrability of $R^n$ models with a scalar or phantom scalar field is an open question.

We consider in detail the $R^2$-gravity model with a scalar field without potential. We obtain the general solutions of evolution equations in the cosmic time and analyze a possible behaviour of the Hubble parameter. Also, we study the correspondence between solutions in the Jordan and Einstein frames and show that it is not one-to-one correspondence for all solutions in the case of a phantom scalar field. In particular, we have found such analytic solutions that the Ricci scalar changes its sign during evolution. These solutions have no analogues in the Einstein frame.

The paper is organized as follows. In Section 2, we define the chiral cosmological model and construct the corresponding modified gravity models. In Section 3, we find the general solution of this chiral cosmological model. In Section 4, we consider the corresponding $R^2$ gravity model with an additional scalar field and obtain its general solution in the cosmic time.
In Section 5, we compare the obtained solutions the Jordan and Einstein frames. Section 6 contains concluding remarks.

\section{Chiral Cosmological Models and Modified Gravity}

Our goal is to find integrable modified gravity cosmological models with an additional scalar field minimally coupled to gravity:
\begin{equation}
\label{actionJ}
\begin{split}
\tilde{S}_{\mathrm{J}} &=\! \int\! {d}^4 x \sqrt{-\tilde{g}}\left[ U(\sigma)\tilde{R} -\frac{\theta_\sigma}{2}\tilde{g}^{\mu \nu}\partial_{\mu}\sigma \partial_{\nu}\sigma
\right. \\
   &\left.{}-\frac{\varepsilon_\psi}{2} \tilde{g}^{\mu \nu} \partial_{\mu}\psi \partial_{\nu}\psi-V_J(\sigma)\right],
\end{split}
\end{equation}
where $U(\sigma)>0$ and $V_J(\sigma)$ are double differentiable functions, the constant $\theta_\sigma$ equals either $\pm 1$ or $0$, whereas $\varepsilon_\psi=\pm 1$. The case of $\theta_\sigma=0$ corresponds to  $F(R)$ gravity models.
As known, modified gravity models (\ref{actionJ}) are connected with chiral cosmological models by the metric transformation.

Let us consider the chiral cosmological model, describing by the following action:
\begin{equation}
\label{SE}
\begin{split}
   S_{\mathrm{E}}&=\int d^4 x \sqrt{-g} \left[\frac{M_\mathrm{Pl}^2}{2} R -\frac{1}{2}g^{\mu \nu}\nabla_{\mu}\phi \nabla_{\nu}\phi \right. \\
   &\left.{}-\frac{\varepsilon_\psi}{2}K(\phi) g^{\mu \nu}\nabla_{\mu}\psi \nabla_{\nu}\psi - V_E(\phi) \right],
\end{split}
\end{equation}
where the $K$ and $V_E$ are differentiable functions. We consider $K(\phi)$ to be a positive-definite function, so the field $\psi$ is either phantom or ordinary scalar field in dependence of $\varepsilon_\psi$. Dynamics of cosmological solutions of models (\ref{SE}) with potentials depending on the field $\phi$ only are actively studied~\cite{Paliathanasis,Paliathanasis:2020sfe,Paliathanasis:2021fxi}.

The considered models with the phantom scalar field $\psi$ are generalizations of quintom models that have been actively used to describe dark energy with the crossing of the cosmological constant barrier (see review~\cite{QuntomREV}). There exist methods for construction of quintom models with exact solutions~\cite{Arefeva:2005mka,Lazkoz:2006pa,Vernov:2006dm,Andrianov:2007ua,Setare2008,Arefeva:2009tkq,Leon:2018lnd} as well as for construction of chiral cosmological models with exact solutions~\cite{Chervon:2019nwq,Paliathanasis:2014yfa}, but it is a almost unsolvable problem to construct integrable cosmological models with multiple scalar fields~\cite{Chimento:1998ju,Leon:2018lnd}.

Using the conformal transformation of the metric:
\begin{equation}
\label{metrictrans}
    g^{\mu \nu}=K(\phi)\tilde{g}^{\mu \nu},
\end{equation}
we get the following action in the Jordan frame:
\begin{equation}
\label{actionJ1}
\begin{split}
S_J &=\! \int\! {d}^4 x \sqrt{-\tilde{g}} \left[\frac{M_\mathrm{Pl}^2}{2 K} \tilde{R}-\frac{\tilde{g}^{\mu \nu}}{2 K}\left[1 - \frac{3 M_\mathrm{Pl}^2K_{,\phi}^2}{2K^2}\right]\partial_{\mu}\phi \partial_{\nu}\phi \right. \\
   &\left.{}-\frac{\varepsilon_\psi}{2} \tilde{g}^{\mu \nu} \partial_{\mu}\psi \partial_{\nu}\psi - \frac{V_E}{K^2} \right].
\end{split}
\end{equation}

If
\begin{equation*}
    2K^2 > 3 M_\mathrm{Pl}^2K_{,\phi}^2,
\end{equation*}
then we introduce $\sigma$ by the relation
\begin{equation}\label{sigmaphi}
    \frac{d\sigma}{d\phi}=\sqrt{\frac{1}{K}\left(1 - \frac{3 M_\mathrm{Pl}^2K_{,\phi}^2}{2K^2}\right)}
\end{equation}
in order to get action (\ref{actionJ}) with  $\theta_\sigma=1$,
\begin{equation}
U(\sigma)=\frac{M_\mathrm{Pl}^2}{2 K(\phi(\sigma))},\qquad V_J(\sigma)=\frac{V_E(\phi(\sigma))}{K^2(\phi(\sigma))}\,.
\end{equation}

If
\begin{equation*}
    2K^2 < 3 M_\mathrm{Pl}^2K_{,\phi}^2,
\end{equation*}
then a phantom scalar field in the Jordan frame corresponds to standard scalar field $\phi$ in the Einstein frame.
At $2K^2 = 3 M_\mathrm{Pl}^2K_{,\phi}^2$ for all $\phi$, we get $F(R)$ gravity model.

We also consider the case of the exponential function $K(\phi)$:
\begin{equation}
\label{Kphi}
K(\phi)=K_0 \mathrm{e}^{\kappa\phi},
\end{equation}
where $K_0>0$ and $\kappa$ are constants. Note that such chiral cosmological models are actively studied~\cite{DiMarco:2002eb,Braglia:2020eai,Paliathanasis:2020sfe,Paliathanasis:2021fxi}.
To get the general solutions for the Einstein frame model in analytic form we restrict ourselves to the case of a constant potential $V_E(\phi)=\Lambda$.

If $\kappa^2<2/(3M_\mathrm{Pl}^2)$,  we use Eq.~(\ref{sigmaphi}):
\begin{equation}
\frac{d\sigma}{d\phi}=\sqrt{\frac{2-3M_\mathrm{Pl}^2\kappa^2}{2K_0}}\,\mathrm{e}^{\kappa\phi/2},
\end{equation}
and obtain
\begin{equation}
\sigma=\sqrt{\frac{C_\kappa}{K_0}}\,\mathrm{e}^{\kappa\phi/2},
\end{equation}
where
\begin{equation}
\label{Ckappa}
     C_\kappa=\frac{2\left[2-3M_\mathrm{Pl}^2\kappa^2\right]}{\kappa^2}\,.
\end{equation}

Action (\ref{actionJ}) takes the following form:
\begin{equation}
\label{actionInduced}
\begin{split}
\tilde{S}_J &=\! \int\! {d}^4 x \sqrt{-\tilde{g}} \left[\frac{M_\mathrm{Pl}^2}{2C_\kappa}\sigma^2\tilde{R} -\frac{\tilde{g}^{\mu \nu}}{2}\nabla_{\mu}\sigma \nabla_{\nu}\sigma\right. \\
   &\left.{}-\frac{\varepsilon_\psi}{2} \tilde{g}^{\mu \nu} \nabla_{\mu}\psi \nabla_{\nu}\psi - \Lambda C_\kappa^2\sigma^4\right],
\end{split}
\end{equation}

In terms of $\sigma$, we obtain an induced gravity model with the fourth degree potential and the additional scalar field~$\psi$.

If
\begin{equation}
\kappa=\pm\frac{\sqrt{2}}{\sqrt{3}M_\mathrm{Pl}}\equiv\kappa_1,
\end{equation}
then the scalar field $\phi$ has no kinetic term in the action $S_J$:
\begin{equation}
\label{SJ_FR_prel}
\begin{split}
S_J &=\! \int\! {d}^4 x \sqrt{-\tilde{g}} \left[\frac{M_\mathrm{Pl}^2}{2 K_0} \mathrm{e}^{-\kappa_1 \phi} \tilde{R}\right. \\
   &\left.{}- \frac{\varepsilon_\psi}{2} \tilde{g}^{\mu \nu} \nabla_{\mu}\psi \nabla_{\nu}\psi - \frac{\Lambda}{K_0^2} \mathrm{e}^{-2 \kappa_1 \phi} \right].
\end{split}
\end{equation}
Varying action (\ref{SJ_FR_prel}) over $\phi$, we obtain for $\Lambda\neq 0$:
\begin{equation}
\label{phiR}
\mathrm{e}^{-\kappa_1 \phi}=\frac{M_\mathrm{Pl}^2K_0}{4\Lambda}\tilde{R}.
\end{equation}

So, we obtain the following action of $F(R)$ gravity with the scalar field $\psi$:
\begin{equation}
\label{actionFR}
S_F = \! \int\! {d}^4 x \sqrt{-\tilde{g}} \left[\frac{M_\mathrm{Pl}^4}{16\Lambda}\,\tilde{R}^2 - \frac{\varepsilon_\psi}{2} \tilde{g}^{\mu \nu} \nabla_{\mu}\psi \nabla_{\nu}\psi \right].
\end{equation}

As known~\cite{Starobinsky:2007hu}, a $F(R)$ gravity model is stable if $F'>0$ and $F''>0$, so, the model $S_F$ is stable for $\tilde{R}>0$. The stable model corresponds to $\Lambda>0$.

In the case of $\kappa^2>2/(3M_\mathrm{Pl}^2)$,  we  introduce a new phantom scalar field $\zeta$ using  relation~(\ref{sigmaphi}):
\begin{equation}
\zeta=\sqrt{\frac{2(3M_\mathrm{Pl}^2\kappa^2-2)}{K_0\kappa^2}}\,\mathrm{e}^{\kappa\phi/2},
\end{equation}
and get
\begin{equation}
\label{actionInducedPhantom}
\begin{split}
S_{ph} &=\! \int\! {d}^4 x \sqrt{-\tilde{g}} \left[{}-\frac{M_\mathrm{Pl}^2}{2}C_\kappa\zeta^2\tilde{R} +\frac{\tilde{g}^{\mu \nu}}{2}\nabla_{\mu}\zeta\nabla_{\nu}\zeta \right. \\
   &\left.{}-\frac{\varepsilon_\psi}{2} \tilde{g}^{\mu \nu} \nabla_{\mu}\psi \nabla_{\nu}\psi - \Lambda C_\kappa^2\zeta^4\right].
\end{split}
\end{equation}
Note that $C_\kappa<0$ in this case. The corresponding one-field cosmological model has been considered in~\cite{Bamba:2014kza}.

The main goal of our paper is to find general solutions of the modified gravity cosmological models described by action (\ref{actionJ}) in the spatially flat Friedmann--Lema\^{i}tre--Ro\-bertson--Walker (FLRW) metric. The standard way to integrate Friedmann equations includes the suitable choice of the parametric time, so  we use
\begin{equation}
ds^2 = {}-N_J^2(\tau)d\tau^2 +a_J^2(\tau)\left(dx_1^2+dx_2^2+dx_3^2\right),
\label{Fried}
\end{equation}
where $a_J(\tau)$ is the scale factor and $N_J(\tau)$ is the lapse function. In this metric, action~(\ref{actionJ}) gives the following equations~\cite{Kamenshchik:2013dga}:
\begin{equation}
6Uh^2+6U'h\dot{\sigma}=\frac{\theta_\sigma}{2}\dot{\sigma}^2+\frac{\varepsilon_\psi}{2}\dot{\psi}^2+N_J^2V_J,
\label{Fried1}
\end{equation}
\begin{equation}
\begin{split}
&4U\dot{h}+6Uh^2+4U'h\dot{\sigma}-4Uh\frac{\dot{N_J}}{N_J}+2U''\dot{\sigma}^2
+2U'\ddot{\sigma}\\
&={} 2U'\dot{\sigma}\frac{\dot{N_J}}{N_J}-\frac {\theta_\sigma}{2}\dot{\sigma}^2-\frac{\varepsilon_\psi}{2}\dot{\psi}^2+N_J^2V_J,
\end{split}
\label{Fried2}
\end{equation}
\begin{equation}
\label{equpsi}
    \ddot{\psi}+\left(3h-\frac{\dot{N_J}}{N_J}\right)\dot{\psi}=0,
\end{equation}
where $h=\dot{a}/a$, dots mean derivatives with respect to time $\tau$ and primes mean derivatives with respect to the scalar field $\sigma$. Note that $h(\tau)$ is the Hubble parameter only for $N_J(\tau)\equiv 1$. The evolution equation in $\sigma$ is a consequence of Eqs.~(\ref{Fried1})--(\ref{equpsi}).

The knowledge of solutions in the Einstein frame allows us to get solutions in the Jordan frame, choosing the parametric time $\tau=t$, where $t$ is the cosmic time in the Einstein frame.
The metric transformation (\ref{metrictrans}) corresponds to the following transformations of the functions defining the FLRW metric:
\begin{equation}
N_J = \sqrt{K(\phi)}N_E,\qquad a_J =  \sqrt{K(\phi)}a_E.
\label{Na}
\end{equation}

So, we get the following solution in the Jordan frame
\begin{equation}
\label{Njt}
N_J(t) =\sqrt{K_0} \mathrm{e}^{\kappa\phi(t)/2},
\end{equation}
\begin{equation}
a_J(t)=\sqrt{K_0} \mathrm{e}^{\kappa\phi(t)/2} a_E(t),
\end{equation}
\begin{equation}\label{sigmat}
    \sigma(t)=\sqrt{\frac{C_\kappa}{K_0}}\,\mathrm{e}^{\kappa\phi(t)/2}
\end{equation}
and $\psi(t)$ is the same in the both frames.

Let us remind that the cosmic time in the Jordan frame is
\begin{equation}
\label{timetime}
\tilde{t} = \int \sqrt{K(\phi(t))}\,dt,
\end{equation}
and the Hubble parameter in the Jordan frame
\begin{equation}
\label{H_J}
H_J(\tilde{t}) = \frac{1}{\sqrt{K(\phi(\tilde{t}))}} \left[H_E(\tilde{t}) + \frac12 \frac{d \ln K}{d t} (\tilde{t})\right].
\end{equation}
So, using Eq.~(\ref{timetime}), we can get general solutions in the cosmic time $\tilde{t}$ for some functions $K(\phi)$ at least in quadratures.
For the $R^2$ model, we consider this question in detail in Section~\ref{EiJo}.

\section{General solutions for Friedmann equations in the Einstein frame}

Varying action (\ref{SE}) with $V_E=\Lambda$, and substituting the spatially flat FLRW metric with
\begin{equation*}
ds^2={}-dt^2+a_E^2(t)\left(dx^2+dy^2+dz^2\right),
\end{equation*}
we obtain the following system of equations:
\begin{equation}
\label{forconstraint}
  3M_\mathrm{Pl}^2H_E^2 = \frac{1}{2}\dot{\phi}^2 + \frac{\varepsilon_\psi}{2}K\dot{\psi}^2 +  \Lambda,
\end{equation}
\begin{equation}
\label{forHdotold}
  2M_\mathrm{Pl}^2 \dot{H}_E + 3M_\mathrm{Pl}^2H_E^2 + \frac{1}{2}\dot{\phi}^2 + \frac{\varepsilon_\psi}{2}K\dot{\psi}^2 =  \Lambda,
\end{equation}
\begin{equation}
\label{forphiddot}
  \ddot{\phi} ={} -3 H_E \dot{\phi} +  \frac{\varepsilon_\psi}{2}K'_{,\phi}\dot{\psi}^2,
\end{equation}
\begin{equation}
\label{forpsiddot}
  \ddot{\psi} = {} - 3 H_E \dot{\psi} - \frac{K'_{,\phi}}{K}\dot{\phi} \dot{\psi},
\end{equation}
where $H_E=\dot{a}_E/a_E$, $a_E(t)$ is the scale factor in the Einstein frame, dots and primes denote the derivatives with respect to the cosmic time $t$ and to the scalar field $\phi$ respectively.

From Eqs.~(\ref{forconstraint}) and (\ref{forHdotold}), we get
\begin{equation}
\label{equ10}
\dot{H}_E + 3H_E^2  = \lambda,
\end{equation}
where $\lambda\equiv \Lambda/M_\mathrm{Pl}^2$.

If $\Lambda>0$, then the general solution of Eq.~(\ref{equ10}) is
\begin{equation}
\label{w0_Hsol}
H_E(t) = \sqrt{\frac{\lambda}{3\,}}\,\frac{1 - C \mathrm{e}^{-2\sqrt{3\lambda}\,t}}{1 + C \mathrm{e}^{-2\sqrt{3\lambda}\,t}},
\end{equation}
where  $C$ is an integration constant.

The corresponding scale factor is
\begin{equation}\label{at}
   a_E(t)=a_0\mathrm{e}^{\sqrt{\frac{\lambda}{3}}\,t}\left(1+C\mathrm{e}^{-2\sqrt{3\lambda}\,t}\right)^{1/3}\,,
\end{equation}
where $a_0$ is a constant.

We get three forms of the Hubble parameter in dependance of the sign of $C$:
\begin{equation}
\label{w0_Hsoltanh}
H_E(t) = \sqrt{\frac{\lambda}{3}}\,\tanh\left(\sqrt{3\lambda}\,(t-t_0)\right),
\end{equation}
for $C>0$,
\begin{equation}
\label{HEdS}
H_E= \sqrt{\frac{\lambda}{3}}
\end{equation}
for $C=0$, and
\begin{equation}\label{w0_Hsolcoth}
  H_E(t) = \sqrt{\frac{\lambda}{3}}\,\coth\left(\sqrt{3\lambda}\,(t-t_0)\right),
\end{equation}
for $C<0$.

To get $\phi(t)$ for the exponential function $K(\phi)$ we pre\-sent Eq.~(\ref{forphiddot}) in the following form:
\begin{equation}
\label{phi_w0}
\begin{split}
  \ddot{\phi} &= {}-3 H_E \dot{\phi} + \frac{K'_{,\phi}}{K}\left(3M_\mathrm{Pl}^2H_E^2 - \frac{1}{2}\dot{\phi}^2 - \Lambda\right)\\
&={} -3 H_E \dot{\phi} + 3\kappa M_\mathrm{Pl}^2 H_E^2 - \frac{\kappa}{2} \dot{\phi}^2 - \kappa M_\mathrm{Pl}^2 \lambda.
\end{split}
\end{equation}

Introducing a new function
\begin{equation*}
u(t) =\sqrt{\frac{K(\phi)}{K_0}}= \mathrm{e}^{\kappa \phi/2},
\end{equation*}
 we transform Eq.~(\ref{phi_w0}) into the following second-order linear differential equation:
\begin{equation}
\label{equu2}
\ddot{u} + 3 H_E \dot{u} + \frac{\kappa^2 M_\mathrm{Pl}^2}{2}\left(\lambda - 3 H_E^2\right) u = 0.
\end{equation}

Now we introduce a new independent variable, $\chi = \sqrt{3 / \lambda}\,H_E(t)$, that is proportional to the Hubble parameter.
This change of variables gives the following Chebyshev's differential equation:
\begin{equation*}
\left(1 - \chi^2\right) \frac{d^2 u}{d \chi^2} - \chi \frac{d u}{d \chi} + \frac{\kappa^2 M^2_\mathrm{Pl}}{6} u = 0.
\end{equation*}
The general solution of Eq.~(\ref{equu2}) can be presented in the following form:
\begin{equation}
\label{solu}
u(t) = A \cos\left[\frac{\kappa M_\mathrm{Pl}}{\sqrt{6}} \arccos \left[\frac{1 - C \mathrm{e}^{-2\sqrt{3\lambda}\,t}}{1 + C \mathrm{e}^{-2\sqrt{3\lambda}\,t}}\right] + B\right]\!,
\end{equation}
where $A$ and $B$ are constants of integration.

It is more convenient to write the function $\phi$:
\begin{equation*}
%\label{phi_exp}
\phi = \frac{2}{\kappa}\ln\left[A \cos\left(\frac{\kappa M_\mathrm{Pl}}{\sqrt{6}} \arccos \left(\frac{1 - C \mathrm{e}^{-2\sqrt{3\lambda}\,t}}{1 + C \mathrm{e}^{-2\sqrt{3\lambda}\,t}}\right) + B\right)\right],
\end{equation*}
in explicitly real forms for $C>0$ and $C<0$ separately. Namely,  we get
\begin{equation*}
\phi=\frac{2}{\kappa}\ln\!\left[A \cos\left[\frac{\kappa M_\mathrm{Pl}}{\sqrt{6}}\arccos\!\left[\tanh \left(\sqrt{3 \lambda}(t - t_0)\right)\right] + B\right]\right]\!
\end{equation*}
for $C > 0$, and
\begin{equation*}
\begin{split}
\phi &= \frac{2}{\kappa}\ln\left[A \tanh^n \left(\frac{\sqrt{3 \lambda}}{2} (t - t_0)\right) \right.\\
&\left.{} + B \coth^n \left(\frac{\sqrt{3 \lambda}}{2} (t - t_0)\right)\right],
\end{split}
\end{equation*}
where $n = \kappa M_\mathrm{Pl} / \sqrt{6}\,$, for $C < 0$.

Equation~(\ref{forpsiddot}) can be integrated:
\begin{equation}
\label{dpsi}
\dot{\psi}=\frac{C_\psi}{Ka_E^3}=\frac{\tilde{C}_\psi  \mathrm{e}^{-\sqrt{3\lambda}\, t}}{\left(1+C\mathrm{e}^{-2\sqrt{3\lambda}\, t}\right)K_0 u^2},
\end{equation}
where $\tilde{C}_\psi\equiv C_\psi/a_0^3$ is an integration constant.

In terms of $u$, Eq.~(\ref{forconstraint}) has the following form:
\begin{equation}
\label{forconstraintu}
\begin{split}
&4K_0{\dot{u}}^2\left(C^2\mathrm{e}^{-4\sqrt{3\lambda}t}+1\right)+\\
&\left[\varepsilon_\psi \tilde{C}_\psi^2\kappa^2+8CK_0\left(\kappa^2\Lambda u^2+{\dot{u}}^2\right)\right]\mathrm{e}^{-2\sqrt{3\lambda}t}=0.
\end{split}
\end{equation}

Substituting solution (\ref{solu}) into Eq.~(\ref{forconstraintu}), we get
\begin{equation}
8 A^2 K_0 \Lambda C  + \tilde{C}^2_\psi \varepsilon_\psi = 0.
\end{equation}
We see that the solutions with $C > 0$ exist in the case of $\varepsilon_\psi = -1$ only, whereas the solutions with $C < 0$ exist only at $\varepsilon_\psi = 1$.

The Hubble parameter $H_E$ does not depend on the form of the function $K(\phi)$. Moreover, the Hubble parameter~(\ref{w0_Hsolcoth}) can be obtained in the case of
 the corresponding one-field model without~$\psi$~\cite{Kamenshchik:2016epi}, whereas solution~(\ref{w0_Hsoltanh}) exists in the model with a phantom field $\phi$ and a standard field $\psi$ for an arbitrary positive-definite function $K(\phi)$. For example, for $K=\sin^2(\phi/M_{Pl})$  the Hubble parameter~(\ref{w0_Hsoltanh}) and the explicit form of functions $\phi(t)$ and $\psi(t)$ have been found in Ref.~\cite{Chervon:2014dya}. Note that the same behaviour of the Hubble parameter $H_E$ has been found in the model with a nonlocal scalar field~\cite{Arefeva:2007xdy}.

Equation~(\ref{equ10}) has a constant nonzero solutions $H_E = \pm \sqrt{\lambda / 3}$. In this case, one obtains the following general solution for $\phi$:
\begin{equation*}
\phi(t) = \frac{2}{\kappa} \ln \left| \mathrm{e}^{-3 H_E t} - C_1\right| + C_2,
\end{equation*}
where $C_1$ and $C_2$ are constants of integration.

The function $\dot{\psi}$ is given by
\begin{equation*}
\dot{\psi}(t) = \frac{\tilde{C}_\psi \mathrm{e}^{3 H_E t}}{K_0\mathrm{e}^{\kappa C_2}(C_1\mathrm{e}^{3 H_E t} - 1)^2}.
\end{equation*}

So, the constraint equation~(\ref{forconstraint}):
\begin{equation}
\left(\frac{6 H_E}{\kappa}\right)^2 + \varepsilon_\psi \mathrm{e}^{\kappa C_2} \tilde{C}^2_\psi = 0.
\end{equation}
gives us that for a constant $H_E$, the scalar field $\psi$ should be nonconstant and a phantom one.

If $\Lambda=0$, then Eq.~(\ref{equ10}) has the following solution:
\begin{equation}
\label{w0_HsolLambda0}
H_E(t)=\frac{1}{3(t-t_0)}\,,
\end{equation}
therefore
\begin{equation}
\label{w0_asolLambda0}
a_E(t)=a_0\left(t-t_0\right)^{1/3}\,,\qquad \dot{\psi}=\frac{\tilde{C}_\psi}{K_0\left(t-t_0\right)u^2}\,.
\end{equation}

Solving Eq.~(\ref{equu2}) and substituting into Eq.~(\ref{forconstraint}), we get
\begin{equation*}
u(t)=C_u(t-t_0)^{\frac{\kappa}{\sqrt{6}} M_\mathrm{Pl}}+\frac{3\varepsilon_\psi\tilde{C}_\psi^2}{8K_0^3M_\mathrm{Pl}^2C_u}\left(t-t_0\right)^{{}-\frac{\kappa}{\sqrt{6}} M_\mathrm{Pl}},
%\label{utLambda0}
\end{equation*}
where $C_u$ is an integration constant. So,
\begin{equation*}
\phi=\frac{2}{\kappa}\ln\left[C_u(t-t_0)^{\frac{\kappa}{\sqrt{6}} M_\mathrm{Pl}}+\frac{3\varepsilon_\psi\tilde{C}_\psi^2\left(t-t_0\right)^{{}-\frac{\kappa}{\sqrt{6}}M_\mathrm{Pl}}}{8K_0^3M_\mathrm{Pl}^2C_u}\right].
\end{equation*}

These solutions obtained in the Einstein frame allow us to get solutions in the parametric time in the Jordan frame by using formulae (\ref{Njt})--(\ref{sigmat}).

\section{The integrable $R^2$ model with a scalar field}

\subsection{The general solution in the cosmic time}
The results of the previous section is sufficient to get solutions of modified gravity models in the parametric time $\tau=t$. To obtain solutions in the cosmic time $\tilde{t}$ one can use Eq.~(\ref{timetime}) or try to solve the evolution equations with the cosmic time. The second way can give solutions in a more simple analytic form, also it guarantees that no solution has been lost.
In this section, we obtain the general solution of the considered $R^2$ model in the cosmic time.

The $F(R)$  model has the following evolution equations:
\begin{equation}\label{frequ}
    F'R_{\mu\nu}-\frac{1}{2}g_{\mu\nu}F-(\nabla_{\mu}\nabla_\nu-g_{\mu\nu}\Box)F'=\frac12 T_{\mu\nu},
\end{equation}
where $T_{\mu\nu}$ is the matter stress-energy tensor.

Action (\ref{actionFR}) corresponds to $F=F_0 \tilde{R}^2$ with $F_0=\frac{M_\mathrm{Pl}^4}{16\Lambda}$ and the stress-energy tensor
\begin{equation}
\label{Tmunupsi}
    T_{\mu\nu}=\varepsilon_\psi\partial_\mu\psi\partial_\nu\psi-\frac{\varepsilon_\psi}{2}\tilde{g}_{\mu\nu}g^{\alpha\beta}\partial_\alpha\psi\partial_\beta\psi.
\end{equation}

In the FLRW metric with
\begin{equation*}
ds^2={}-d\tilde{t}^2+\tilde{a}^2(\tilde{t})\left(dx^2+dy^2+dz^2\right),
\end{equation*}
system (\ref{frequ}) has only two independent equations:
\begin{equation}
\label{equFR00}
    F'\tilde{R}_{00}+\frac{1}{2}F+3H_J\dot{{F}'}=\frac14{}\varepsilon_\psi{\dot{\psi}}^2\,,
\end{equation}
\begin{equation}
\label{equFR11}
  F'\tilde{R}_{11}-\frac{\tilde{a}^2}{2}F-\tilde{a}^2\left(\ddot{{{F}'}}+2H_J\dot{{F}'}\right)={}\frac{\tilde{a}^2}{4}\varepsilon_\psi{\dot{\psi}}^2\,.
\end{equation}

The trace equation
\begin{equation}\label{equFRtr}
     F'\tilde{R}-2F+3\Box F'=\frac12 T^{\mu}_\mu={}-\varepsilon_\psi{\dot{\psi}}^2.
\end{equation}
is a consequence of Eqs.~(\ref{equFR00}) and (\ref{equFR11}). In this section, dots mean derivatives with respect to the cosmic time $\tilde{t}$.

Using
\begin{equation*}
\begin{split}
    \tilde{R}_{00}&={}-3\left(\dot{H}_J+H_J^2\right),\\ \tilde{R}_{11}&=\tilde{a}^2\left(\dot{H}_J+3H_J^2\right),\\ \tilde{R}&=6\left(\dot{H}_J+2H_J^2\right)\,,
\end{split}
\end{equation*}
we obtain for the $R^2$ model considered:
\begin{equation}
\label{equ00R2}
    18F_0\left(6H_J^2\dot{H}_J-{\dot{H}_J}^2+2H_J\ddot{H}_J\right)=\frac{\varepsilon_\psi}{4}{\dot{\psi}}^2,
\end{equation}
\begin{equation}
\label{equ11R2}
   6F_0\left(18H_J^2\dot{H}_J+12H_J\ddot{H}_J+9{\dot{H}_J}^2+2\dddot{H}_J\right)={}-\frac{\varepsilon_\psi}{4}{\dot{\psi}}^2.
\end{equation}

Excluding $\dot{\psi}$, we get the following third order differential equation in $H_J$:
\begin{equation}
\label{equH}
\dddot{H}_J
+9H_J \ddot{H}_J+18H_J^2\dot{H}_J+3{\dot{H}_J}^2=0.
\end{equation}
Multiplying Eq.~(\ref{equH}) by ${\dot{H}_J}^2$ and factoring, we get
\begin{equation*}
\begin{split}
&\left(\ddot{H}_J + 3 H_J \dot{H}_J\right)\left(2 H_J \dddot{H}_J + 6 H_J^2\ddot{H}_J +12 H_J \dot{H}_J^2\right)\\
&=\left(2 H_J \ddot{H}_J + 6 H_J^2 \dot{H}_J-\dot{H}_J^2\right)\left(\dddot{H}_J + 3 H_J \ddot{H}_J + 3 \dot{H}_J^2\right),
\end{split}
\end{equation*}
or, equivalently,
\begin{equation}
\label{FR_integrals}
\begin{split}
&\left(\ddot{H}_J + 3 H_J \dot{H}_J\right)\,\frac{d}{d\tilde{t}}
\left[\dot{H}_J^2 - 2 H_J \ddot{H}_J - 6 H_J^2 \dot{H}_J\right]\\
&=\left(\dot{H}_J^2 - 2 H_J \ddot{H}_J - 6 H_J^2 \dot{H}_J\right)\,\frac{d}{d\tilde{t}}\left[\ddot{H}_J + 3 H_J \dot{H}_J\right]\!.
\end{split}
\end{equation}

So, there are two families of solutions:
\begin{enumerate}

\item The first possibility is
\begin{equation}\label{equH1}
\ddot{H}_J + 3 H_J \dot{H}_J = 0\,.
\end{equation}

This equation has the following integral:
\begin{equation*}
2\dot{H}_J + 3 H_J^2 = 2\tilde{C},
\end{equation*}
where $\tilde{C}$ is an integration constant.

For this case, Eq.~(\ref{equ00R2}) takes the following form:
\begin{equation}
\varepsilon_\psi{\dot{\psi}}^2={}-72 F_0 {\dot{H}}_J^2.
\end{equation}

From this relation, it follows that $\varepsilon_\psi = -1$, and
\begin{equation}
\dot{\psi} = \pm 6 \sqrt{2 F_0} \dot{H}_J=\pm 3 \sqrt{2 F_0}\left(2\tilde{C}-3 H_J^2\right).
\end{equation}

The model has de Sitter solutions with $H_J=\sqrt{2\tilde{C}}/3$ that correspond to a constant $\psi$.

The type of solutions obtained depends on the sign of $\tilde{C}$, see Table~\ref{T1}. The values of constants $B$ and $\tilde{t'}$ are defined by the initial value $H_{J0}$. The third line of Table~\ref{T1} includes the solutions from lines 1 and 2 in a different form and de Sitter solutions at $B=0$. The power-law solution is presented in the fourth line.
One needs to be careful with the solution from the fifth line, because the scalar curvature $\tilde{R}$ changes sign at $\tilde{t} = \tilde{t'} \pm \frac{\pi}{3}\sqrt{-2/ (3 \tilde{C})}$.

\begin{table}[h]
\begin{center}
\caption{List of the $R^2$ gravity exact solutions.}
\begin{tabular}{|c|c|c|}
\hline
$\tilde{C}$&   $H_J\left(\tilde{t}\right)$ & $\left|{\dot{\psi}}\left(\tilde{t}\right)\right|$ \\[1mm]
\hline
$\begin{array}{c}
\tilde{C}>0,\\
\! \dot{H}_{J0} > 0\!
\end{array}$
&  $\sqrt{\frac{2 \tilde{C}}{3}} \tanh \left(\sqrt{\frac{3 \tilde{C}}{2}} \left(\tilde{t}-\tilde{t'}\right)\right)$ &
$\frac{6 \tilde{C} \sqrt{2 F_0}}{\cosh^2 \left(\sqrt{\frac{3 C}{2}} \left(\tilde{t}-\tilde{t'}\right)\right)}\!$  \\
\hline
$\begin{array}{c}
\tilde{C}>0,\\
\!\dot{H}_{J0} < 0\!
\end{array}$
 &  $\sqrt{\frac{2 \tilde{C}}{3}} \coth \left(\sqrt{\frac{3 \tilde{C}}{2}} \left(\tilde{t}-\tilde{t'}\right)\right)$ & $\frac{6 \tilde{C} \sqrt{2 F_0}}{\sinh^2
 \left(\sqrt{\frac{3 C}{2}}\left(\tilde{t}-\tilde{t'}\right)\right)}\!$ \\
\hline
$\tilde{C}>0$   &  $\frac{\sqrt{6\tilde{C}}\left(1-B\mathrm{e}^{-\sqrt{6\tilde{C}}\tilde{t}}\right)}{3\left(1+B\mathrm{e}^{-\sqrt{6\tilde{C}}\tilde{t}}\right)}$
& $ \frac{24BC\sqrt{2F_0}\mathrm{e}^{-\sqrt{6\tilde{C}}\tilde{t}}}{\left(B\mathrm{e}^{-\sqrt{6\tilde{C}}\tilde{t}}+1\right)^2}\!$\\
\hline
$\tilde{C}=0$ & $\frac{2}{3\left(\tilde{t}-\tilde{t'}\right)}$ & $\frac{4 \sqrt{2 F_0}}{\left(\tilde{t}-\tilde{t'}\right)^2}$ \\
\hline
$\tilde{C}<0$ & $\!-\frac{\sqrt{-6\tilde{C}}}{3}\tan\left[\frac{\sqrt{-6\tilde{C}}}{2} \left(\tilde{t}-\tilde{t'}\right)\right]\!\!$ &
$\frac{6 C \sqrt{2 F_0}}{\cos^2 \left(\sqrt{\frac{-3 C}{2}} \left(\tilde{t}-\tilde{t'}\right)\right)}\!$ \\
\hline
\end{tabular}
\label{T1}
\end{center}
\end{table}

\item
If $\ddot{H}_J + 3 H_J \dot{H}_J \neq 0$, then one can integrate Eq.~(\ref{FR_integrals}) and get the following equation:
\begin{equation}
\frac{\dot{H}_J^2}{\ddot{H}_J + 3 H_J \dot{H}_J}-2H_J = C_1,
\end{equation}
where $C_1$ is a constant of integration.

Integrating the equation,
\begin{equation}
\label{hddot_eqn}
(C_1 + 2 H_J) \ddot{H}_J + 3 H_J (C_1 + 2 H_J) \dot{H}_J - \dot{H}_J^2 = 0,
\end{equation}
one gets:
\begin{equation}
\label{hdot_eqn}
\dot{H}_J = C_2 \sqrt{\left| C_1 + 2 H_J \right|} + (C_1 + 2 H_J) (C_1 - H_J),
\end{equation}
where $C_2$ is also a constant of integration.

Equation (\ref{hdot_eqn}) with arbitrary constants $C_1$ and $C_2$ can be solved in quadratures.
Also, there are some particular solutions of Eq.~(\ref{hdot_eqn}):
\begin{enumerate}
\item At $C_1 = C_2 = 0$,
\begin{equation}
\label{R0sol}
H_J(t) = \frac{1}{2\left(\tilde{t}-\tilde{t'}\right)},
\end{equation}
\item At $C_2 = 0$, $C_1\neq 0$,
\begin{equation}
H_J(t) = C_1 \frac{\tilde{C} + \mathrm{e}^{-3 C_1 \tilde{t}}}{\tilde{C} - 2\, \mathrm{e}^{-3 C_1 \tilde{t}}},
\label{C20sol}
\end{equation}
\end{enumerate}
where $\tilde{C}$ is a constant of integration.

Combining Eqs.~(\ref{equ00R2}) and (\ref{hddot_eqn}), we obtain
\begin{equation}
\label{psiequ2}
    {\dot{\psi}}^2={}-72F_0C_1\varepsilon_\psi\left(\ddot{H}_J+3H_J\dot{H}_J\right).
\end{equation}
So, the case of $C_1=0$ corresponds to $R^2$ without additional scalar field.
\end{enumerate}

\subsection{The different behaviour of the Hubble parameter}

\begin{figure}
\includegraphics[width=0.97\linewidth]{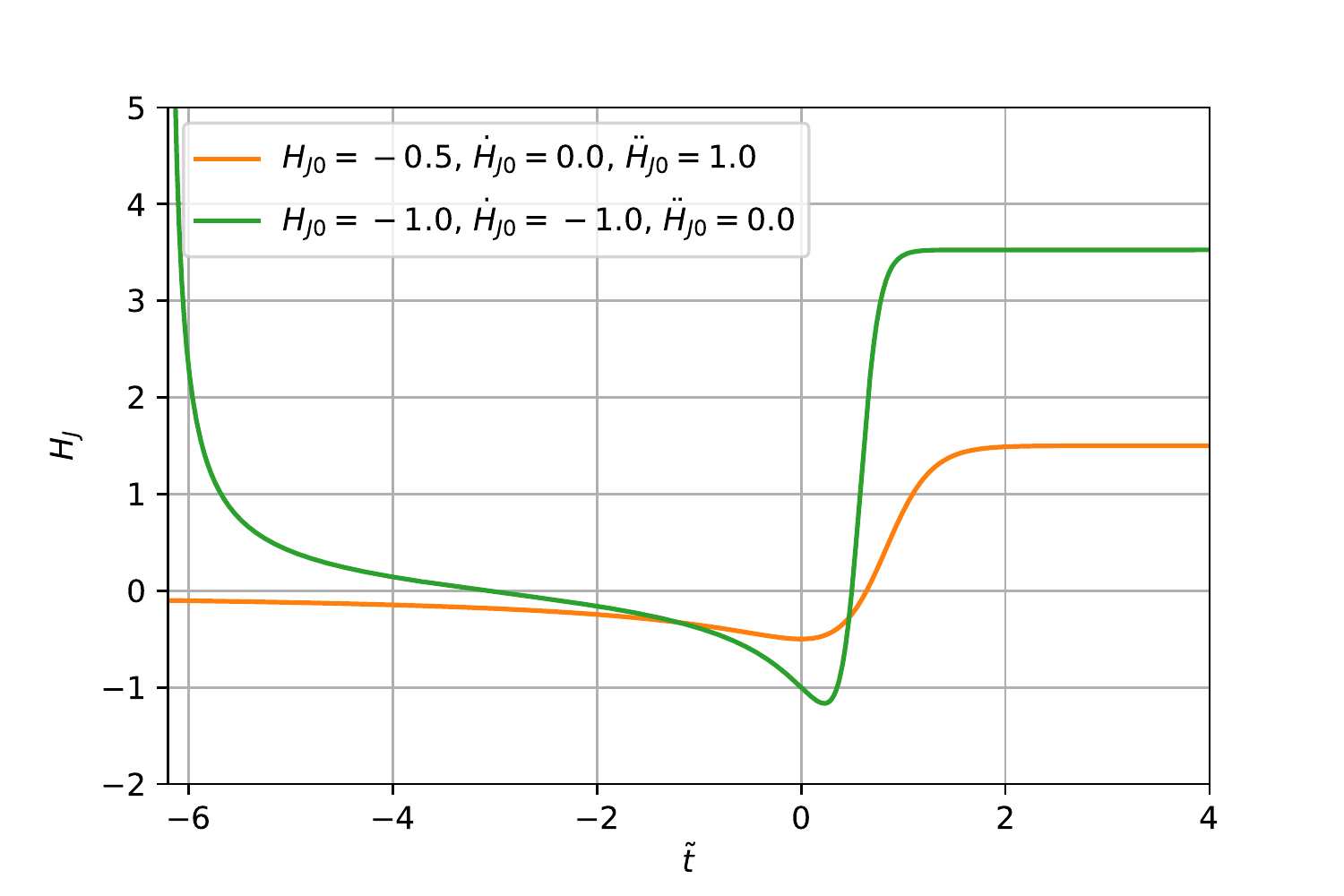}
\caption{The Hubble parameter $H_J(\tilde{t})$ in the $R^2$ model considered.
\label{HJ1}}
\end{figure}

\begin{figure}[h!tbp]
\includegraphics[width=0.49\linewidth]{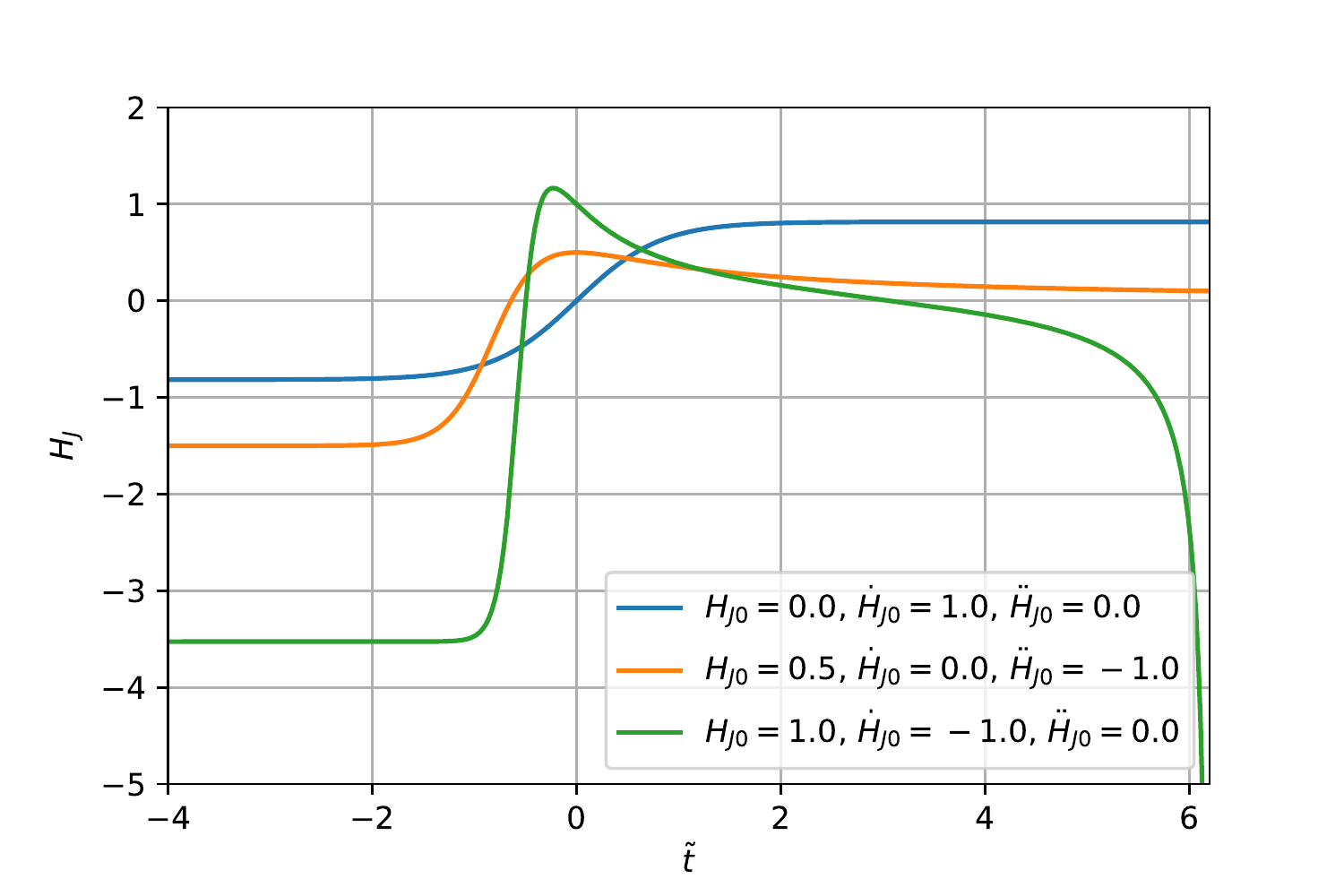} \
\includegraphics[width=0.49\linewidth]{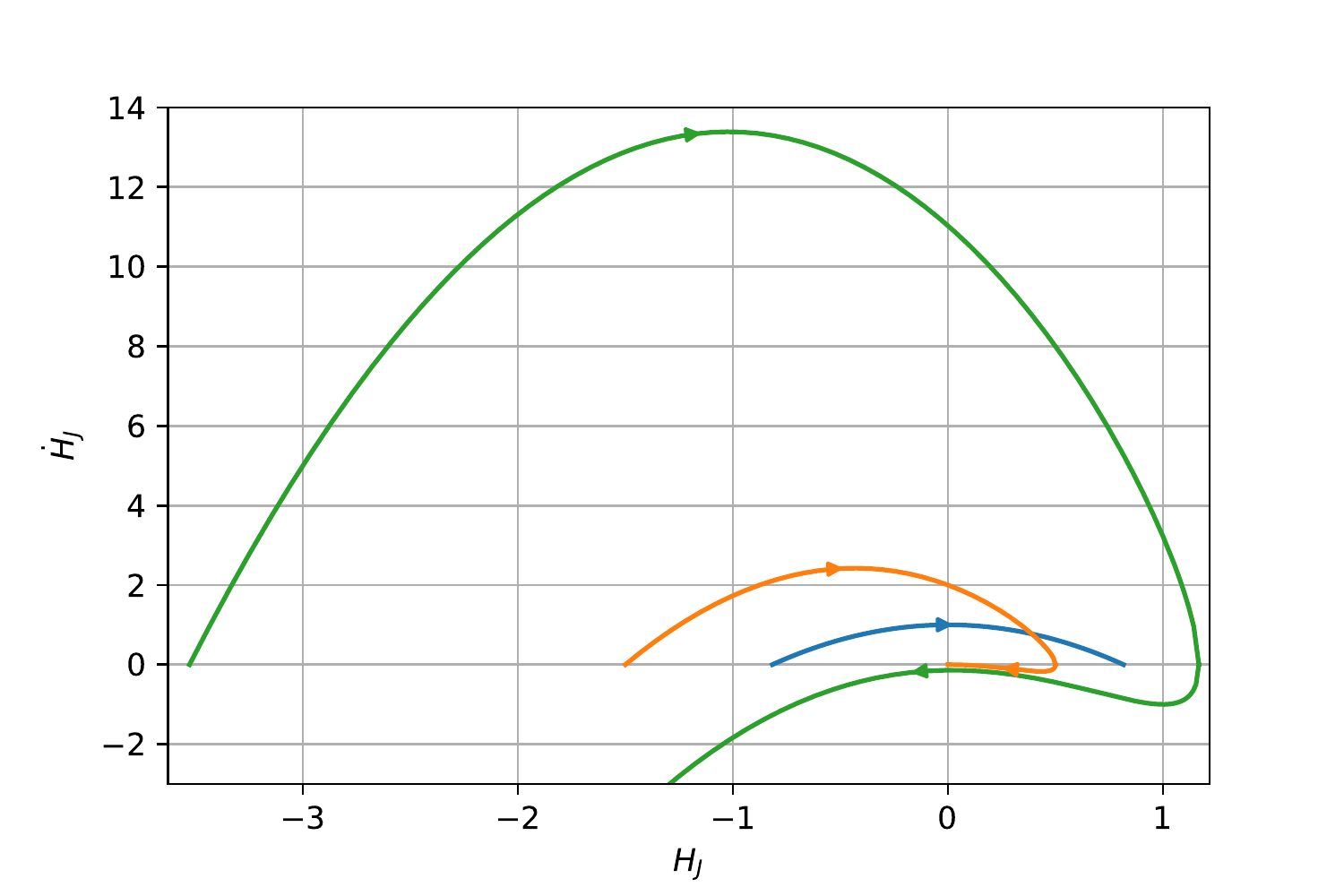}
\caption{The Hubble parameter $H_J(\tilde{t})$ (left) and the corresponding phase portraits (right).
\label{HJ2}}
\end{figure}

One can see in Table~\ref{T1} that all nonconstant solutions of Eq.~(\ref{equH1}) are monotonic functions. Let us consider a possible extremum of a solution of Eq.~(\ref{hdot_eqn}).
Solving Eq.~(\ref{equH}) numerically, we have found solutions with nonmonotonic behaviour. Also, we have found both bounded, and unbounded solutions (see Figs.~\ref{HJ1} and~\ref{HJ2}).
In Fig.~\ref{HJ1}, one can see bounded and unbounded solutions with minima. A monotonic solution and solutions with maxima are presented in Fig.~\ref{HJ2}.


\begin{figure}
\includegraphics[width=0.97\linewidth]{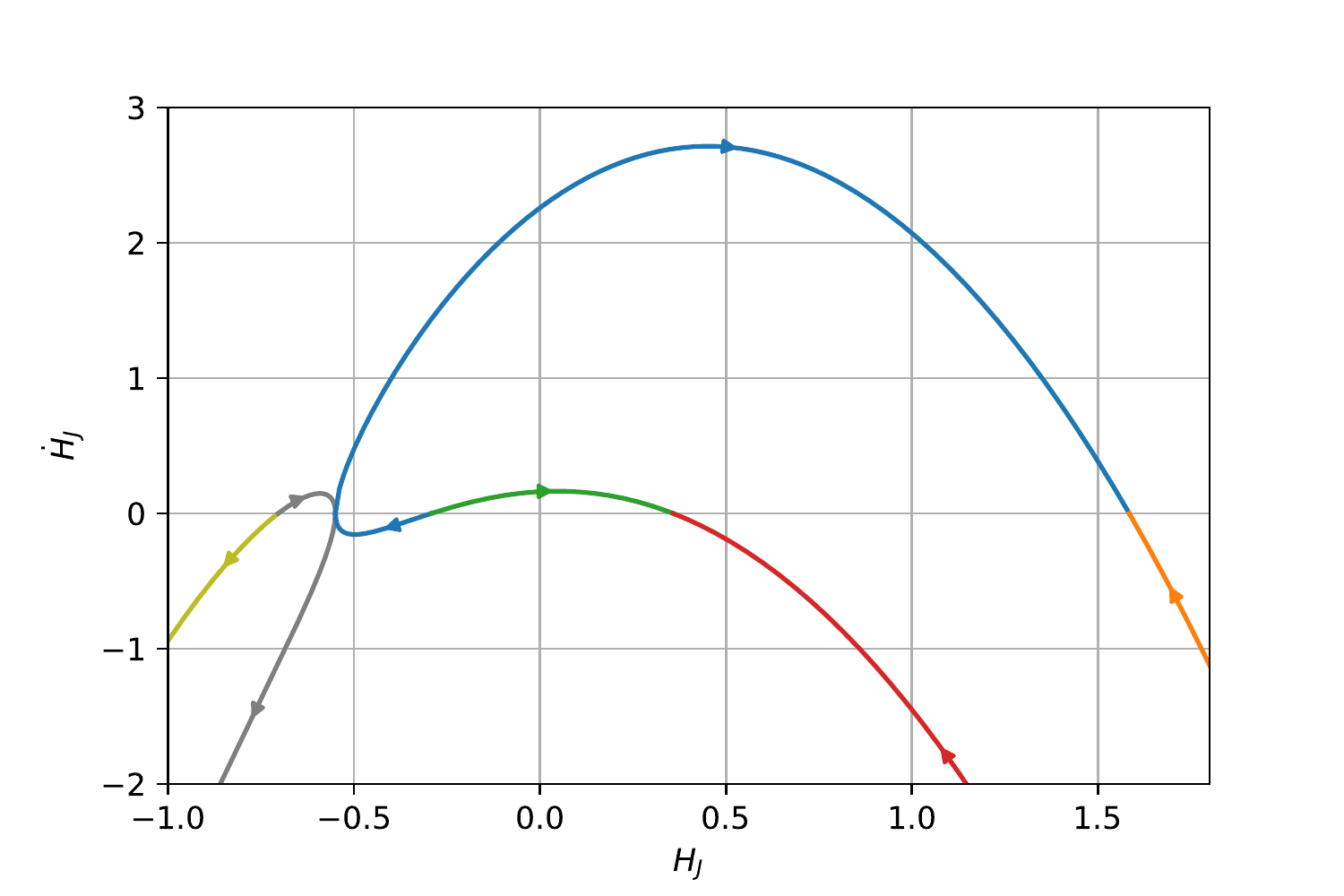}
\caption{Phase portraits $\dot{H}_J(H_J)$ for $C_1 = 1.1$, $C_2 = \pm 1$. Parts of the curves corresponding to the different possible solutions are in different colors. The blue and gray trajectories are described by two equations --- Eq.~(\ref{hdot_eqn}) and Eq.~(\ref{hdot_eqn_after}). The change of the equation describing these trajectories occurs at $H_J = -C_1/2 = -0.55$.
\label{phase_plot_colored}}
\end{figure}

Let us analyze possible types of the Hubble parameter evolution for the phantom and the standard scalar field $\psi$.
At the point $H_J = -C_1/2$, it is obvious that $\dot{H}_J = 0$, but $\ddot{H}_J = C_2^2\,\mathrm{sgn}(C_1 + 2 H_{J0}) \neq 0$ if $C_2 \neq 0$.
This means that $H_J = -C_1/2$ is an extreme value of $H_J$, namely, a maximum for $C_1 + 2 H_{J0} < 0$ and a minimum for $C_1 + 2 H_{J0} > 0$, where $H_{J0} = H_J(\tilde{t}_0)$ is an initial condition for $H_J$. When (if at all), during its evolution, $H_J$ becomes equal to $-C_1/2$, the equation describing $\dot{H}_J$ changes
from~(\ref{hdot_eqn}) to
\begin{equation}
\label{hdot_eqn_after}
\dot{H}_J = -C_2 \sqrt{\left| C_1 + 2 H_J \right|} + (C_1 + 2 H_J) (C_1 - H_J)\,.
\end{equation}
The phase portraits in Fig.~\ref{phase_plot_colored} illustrate this situation.

Combining Eqs.~(\ref{equ00R2}),~(\ref{hddot_eqn}), and~(\ref{hdot_eqn}), one gets the following relation:
\begin{equation*}
-\varepsilon_H F_0 C_1\left[\pm  C_2 + \varepsilon_H\sqrt{|C_1 + 2 H_J|}(C_1 - H_J)\right]^2 = \frac{\varepsilon_\psi}{72}{\dot{\psi}}^2,
\end{equation*}
where $\varepsilon_H = \mathrm{sgn}(C_1 + 2 H_{J0})$, and the symbol ``$\pm $'' before $C_2$ represents the change of equation describing $\dot{H}_J$ when $H$ reaches the value $-C_1/2$. From here, it follows that $\varepsilon_\psi = -\varepsilon_H\,\mathrm{sgn}(C_1)$, and
\begin{equation*}
\dot{\psi} = \pm 6 \sqrt{2 F_0 |C_1|}\left|\pm  C_2 + \varepsilon_H\sqrt{|C_1 + 2 H_J|}(C_1 - H_J)\right|.
\end{equation*}
In particular, if $C_1 = 0$, then $\dot{\psi} \equiv 0$.

Let us find roots of Eqs.~(\ref{hdot_eqn}) and~(\ref{hdot_eqn_after}) that are not equal to $H_J = -C_1/2$. These roots are of interest to us because they correspond to the stable points (asymptotically stable points, to be precise) of Eq.~(\ref{equH}). Knowing how many stable points there are, we can determine for any particular values of $C_1$ and $C_2$ whether or not there exists a bounded solution. At $C_2=0$, solutions are given by Eqs.~(\ref{R0sol}) and (\ref{C20sol}) in the analytic form, so, we need to consider only the case of $C_2\neq 0$.

We get two cases, $2 H_{J0} + C_1 > 0$ and $2 H_{J0} + C_1 < 0$. We start by analyzing the first case. The substitution $\tilde{u} = \sqrt{2 H_J + C_1}$ results in the following pair of equations:
\begin{equation*}
\pm C_2 + \frac12 \tilde{u} \left(3 C_1 - \tilde{u}^2\right) = 0 \iff \tilde{u}^3 - 3 C_1 \tilde{u} \pm 2 C_2 = 0.
\end{equation*}

The determinant of both of these equations is
\begin{equation}
\label{determ_1st}
\Delta = 108 \left(C_1^3 - C_2^2\right).
\end{equation}
So, for $C_1^3 < C_2^2$, we have two real roots:
\begin{equation}
\label{u2}
\tilde{u} = \pm \left[\sqrt[3]{C_2 + \sqrt{C_2^2 - C_1^3}} + \sqrt[3]{C_2 - \sqrt{C_2^2 - C_1^3}}\right],
\end{equation}
and for $C_1^3 > C_2^2$, we have six real roots:
\begin{equation}
\label{u6}
\tilde{u}_k = {} \pm 2 \sqrt{C_1} \cos\left(\frac13 \arccos\left(\frac{C_2}{C_1}\sqrt{\frac{1}{C_1}}\right) - \frac{2 \pi k}{3}\right),
\end{equation}
here and in the following, $ k = 0, 1, 2$.

At  $C_1^3 = C_2^2$, we can use Eq.~(\ref{u6}) as well and get four different roots:
\begin{equation}
\label{u2eq}
\begin{split}
\tilde{u}_k &= \pm 2\sqrt{C_1}\cos\left(\frac{2 \pi k}{3}\right), { \ } \Rightarrow { \ }\\
 \tilde{u}_0 &= {}\pm2\sqrt{C_1},\qquad \tilde{u}_1 = {}\pm\sqrt{C_1}.
 \end{split}
\end{equation}

In terms of $H_J = (\tilde{u}^2 - C_1)/2$, we get:
At $C_1^3 < C_2^2$:
\begin{equation*}
\begin{split}
H_{Jas} &= \frac12 \left[\sqrt[3]{C_2 + \sqrt{C_2^2 - C_1^3}} + \sqrt[3]{C_2 - \sqrt{C_2^2 - C_1^3}}\right]^2\\
&{} - \frac{C_1}{2},
\end{split}
\end{equation*}
at $C_1^3 > C_2^2$:
\begin{equation*}
H_{Jas} = 2 C_1 \cos^2\left(\frac13 \arccos\left(\frac{C_2}{C_1^{3/2}}\right) - \frac{2 \pi k}{3}\right) - \frac{C_1}{2}.
\end{equation*}
and at $C_1^3 = C_2^2$:
\begin{equation*}
H_{Jas} = \frac{3}{2} C_1,\quad \mbox{and}\quad H_{Jas} =0.
\end{equation*}

In a similar manner, one can obtain the values corresponding to the stable points in the case of $2 H_{J0} + C_1 < 0$:
\begin{itemize}
\item at $C_1^3 > -C_2^2$:
\begin{equation*}
\begin{split}
H_{Jas}=&
\frac{-1}{2}\! \left[\sqrt[3]{C_2 + \sqrt{C_2^2 + C_1^3}} + \sqrt[3]{C_2 - \sqrt{C_2^2 + C_1^3}}\right]^2\! \\
&{}- \frac{C_1}{2},
\end{split}
\end{equation*}
\item at $C_1^3 < -C_2^2$:
\begin{equation*}
H_{Jas} = 2 C_1 \cos^2\left[\frac13 \arccos\left(\frac{C_2}{(-C_1)^{3/2}}\right) - \frac{2 \pi k}{3}\right] - \frac{C_1}{2}.
\end{equation*}

\item and at $C_1^3 =-C_2^2$:
\begin{equation*}
H_{Jas} = \frac{3}{2} C_1, \quad \mbox{and}\quad H_{Jas} =0.
\end{equation*}
\end{itemize}

As we mentioned above,
\begin{equation*}
\varepsilon_\psi ={}-\mathrm{sgn}\left(C_1 (C_1 + 2 H_{J0})\right).
\end{equation*}
So, the field $\psi$ is not a phantom field only if $C_1$ and $C_1 + 2 H_{J0}$ are of different signs. But our analysis of the stable points of Eq.~(\ref{equH}) shows that in this case there is only one stable point, and so $H_J(\tilde{t})$ is unbounded. So, we come to the conclusion that bounded solutions are allowed only if the field $\psi$ is a phantom field.

It is easy to verify that continuous and bounded solutions $H_J(\tilde{t})$ of Eqs. (\ref{hdot_eqn}) and (\ref{hdot_eqn_after}) have different asymptotic values at $\tilde{t} \rightarrow \infty$ and $\tilde{t} \rightarrow -\infty$. Obviously, if the asymptotic values of the non-constant solution $H_J(\tilde{t})$ are the same, then the solution has at least one extremum. In this case, $C_2 \neq 0$ and the following equation must have a solution other than $H_J = -C_1/2$:
\begin{equation}
\label{intersect_eqn}
\begin{split}
 &C_2 \sqrt{\left| C_1 + 2 H_J \right|} + (C_1 + 2 H_J) (C_1 - H_J)\\
 &={}-C_2 \sqrt{\left| C_1 + 2 H_J \right|} + (C_1 + 2 H_J) (C_1 - H_J).
\end{split}
\end{equation}
Obviously, it doesn't have any solutions besides $H_J = -C_1/2$. Thus, the asymptotic values at $\tilde{t} \rightarrow \infty$ and $\tilde{t} \rightarrow -\infty$ of the continuous and bounded solution $H_J(\tilde{t})$ of Eqs.~(\ref{hdot_eqn}) and (\ref{hdot_eqn_after}) are never equal.

Another interesting property of the obtained solutions: a continuous solution $H_J(\tilde{t})$ can change sign only if the field $\psi$ is a phantom field. It follows from the fact that
\begin{equation*}
\varepsilon_\psi ={}-\mathrm{sgn}\left(C_1 (C_1 + 2 H_J)\right)\big|_{ H_J = 0} = {}-\mathrm{sgn}(C_1^2) = -1.
\end{equation*}

We can safely take $C_1$ to be nonzero, because the case of $C_1 = 0$ corresponds to $\dot{\psi} \equiv 0$. We obtain that $H_J=0$ at some moment of time only if the field $\psi$ is a phantom one. Therefore, a continuous solution $H_J(\tilde{t})$ can change sign only if the field $\psi$ is a phantom field.

The same result can be obtained by substituting $H_J=0$ into Eq.~(\ref{equ00R2}). Assuming $H_J>0$, Eq.~(\ref{equ00R2}) with $\varepsilon_\psi=1$ can be presented in the following form:
\begin{equation}
\label{equ00R2m}
    \frac{d}{d\tilde{t}}\left[\frac{\tilde{R}}{\sqrt{H_J}}\right]=\frac{{\dot{\psi}}^2}{24 F_0 H_J^{3/2}}\,.
\end{equation}
So, if at the initial moment $\tilde{t}_0$ we have $\tilde{R}(\tilde{t}_0)>0$ and $H_J(\tilde{t}_0)>0$, then $\tilde{R}(\tilde{t})>0$ and $H_J(\tilde{t})>0$ for all $\tilde{t}>\tilde{t}_0$. If the field $\psi$ is a phantom field, the situation is different and the initially positive $\tilde{R}(\tilde{t})$ and $H_J(\tilde{t})$ can change sign during their evolution (see line 5 in Table~\ref{T1} and the green curves in Figs.~\ref{HJ1} and \ref{HJ2}).

\section{The connection between the Jordan and Einstein frame solutions of the $R^2$ model}
\label{EiJo}

In this section, we compare the behaviour of the Hubble parameters in the Jordan and Einstein frame. In the Einstein frame, equations on the Hubble parameter and scalar fields can be separated and the Hubble parameter $H_E$  satisfies the first order differential equation (\ref{equ10}). In the Jordan frame, the Hubble parameter $H_J$ satisfies the third order differential equation (\ref{equH}) and, therefore, includes three free parameters.

For all solutions in the Einstein frame, one can construct the corresponding solutions in the Jordan frame, but some some solutions in Jordan frame have no analogues in the Einstein frame.
Equation~(\ref{phiR}) is sensible and a continuously map from the Jordan frame to the Einstein frame is possible, only if the Ricci scalar $\tilde{R}$ has the same sign as $\Lambda / K_0$ and does not change it during evolution. It is not always the case if the field $\psi$ is a phantom one. For example, the solution for $R^2$ gravity model with $\tilde{C}<0$  (see Table~\ref{T1}) corresponds to $\tilde{R}$ that changes sign. Looking at Figs.~\ref{HJ1} and~\ref{HJ2}, we see that both green curves correspond to solutions with $\tilde{R}$
 that changes sign, because $H_J(\tilde{t})=0$ at two points. Also, solution (\ref{R0sol}) corresponds to $\tilde{R}\equiv 0$. All these solutions cannot be obtained from the Einstein frame solutions.

We consider the case of $\Lambda / K_0>0$. To get the scalar field $\phi(t)$ that corresponds to the given $H_J(\tilde{t})$ we use
\begin{equation*}
6\left(\dot{H}_J(\tilde{t}) + 2 H_J^2(\tilde{t})\right) = \tilde{R}(\tilde{t}) = \frac{4 \Lambda}{K_0 M^2_\mathrm{Pl}} \mathrm{e}^{\sqrt{2/3}\phi(\tilde{t}) /M_\mathrm{Pl}}
\end{equation*}
and
\begin{equation*}
t = \int \frac{\mathrm{e}^{\phi(\tilde{t}) / \sqrt{6} M_\mathrm{Pl}}}{\sqrt{K_0}} d\tilde{t} = \int \sqrt{\frac{3 M^2_\mathrm{Pl}}{2 \Lambda}} \sqrt{\dot{H}_J(\tilde{t}) + 2 H_J^2(\tilde{t})}\,d\tilde{t}.
\end{equation*}

For solutions that correspond to $\tilde{R}>0$, we have
\begin{equation*}
\phi(t(\tilde{t})) = \sqrt{\frac32}  M_\mathrm{Pl} \ln \left[\frac{6 K_0  M^2_\mathrm{Pl}}{4 \Lambda} \left(\dot{H}_J(\tilde{t}) + 2 H_J^2(\tilde{t})\right)\right]
\end{equation*}
and
\begin{equation*}
\frac{d \phi}{d t}\left(t(\tilde{t})\right) = \frac{d \phi}{d \tilde{t}} \frac{d \tilde{t}}{d t} = \sqrt{\Lambda} \frac{\ddot{H}_J + 4 H_J \dot{H}_J}{\left(\dot{H}_J + 2 H^2_J\right)^{3/2}}\,.
\end{equation*}

In our case, Eq.~(\ref{H_J}) takes the form
\begin{equation*}
\begin{split}
H_J(\tilde{t}) &= \frac{\mathrm{e}^{\phi / \sqrt{6} M_\mathrm{Pl}}}{\sqrt{K_0}}\left[H_E(t(\tilde{t})) - \frac{1}{\sqrt{6} M_\mathrm{Pl}} \frac{d \phi}{d t}\left(t(\tilde{t})\right)\right]\\
&= \frac{1}{u(t(\tilde{t}))} \left[H_E(t(\tilde{t})) + \frac{\dot{u}(t(\tilde{t}))}{u(t(\tilde{t}))}\right].
\end{split}
\end{equation*}

Expressing $H_E$ from this equation, we get the following relation:
\begin{equation}
\label{H_J_to_H_E}
\begin{split}
H_E(t(\tilde{t})) &= \sqrt{K_0}\mathrm{e}^{-\phi / \sqrt{6} M_\mathrm{Pl}} H_J(\tilde{t}) + \frac{1}{\sqrt{6} M_\mathrm{Pl}}  \frac{d \phi}{d t}(\tilde{t})
\\&= \sqrt{\frac{\Lambda}{6  M^2_\mathrm{Pl}}}
\frac{\ddot{H}_J + 6 H_J \dot{H}_J + 4 H_J^3}{\left(\dot{H}_J + 2 H_J^2\right)^{3/2}}.
\end{split}
\end{equation}

\section{Conclusion}

In our paper, we have found general solutions in the cosmic time for the two-field chiral cosmological model.
The cosmic time in the Einstein frame corresponds to a parametric time in the Jordan frame, so, the consideration of integrable chiral cosmological models is useful to get the general solutions for the corresponding modified gravity models in parametric time. Using this method, we have found general solutions for the induced gravity cosmological model with the four-order potential and the $R^2$ gravity model. The proposed method allows finding general solutions for different cosmological models that can be presented as chiral cosmological models after the metric transformation. In particular, it would be interesting to generalize our consideration on other forms of the function $K(\phi)$, including $K=\sin^2(\phi/M_{Pl})$ proposed in Ref.~\cite{Chervon:2014dya}. Note that the behaviour of the Hubble parameter (\ref{w0_Hsol}) does not depend on the form of the $K(\phi)$, whereas the form of scalar fields depends on it.

At the same time, some particular solutions can be lost, because they have no analogue in the Einstein frame. For example, constructing the Einstein frame analogue of $R^2$ gravity models, one uses Eq.~(\ref{phiR}), hence, no finite value of $\phi$ corresponds to $\tilde{R}=0$. At the same time, the Ricci scalar $\tilde{R}$ can change its sign during the evolution. We have found such solutions explicitly for the $R^2$ model with a phantom scalar field. On the other hand, it is known that $F(R)$ gravity models without additional scalar field have anisotropic instabilities associated with the crossing of the hypersurface $F'(R)=0$. In other words, the solutions in the FLRW metric are smooth, whereas solutions in the Bianchi I metric have singularities~\cite{Figueiro:2009mm}. A similar situation arises in the model with nonminimally coupled scalar field~\cite{Starobinsky:81} (see also~\cite{Sami:2012uh,Kamenshchik:2017ous,Kamenshchik:2017ojc}).
Anisotropic cosmological solutions in $R + R^2$ gravity have been investigated in~\cite{Muller:2017nxg}.
We plan to analyze anisotropic solutions and their stability in the considering $R^2$ model with an additional scalar field in future investigations.

We have found the general solution of the $R^2$ gravity model with an additional scalar field and analyzed the behaviour of the Hubble parameter.
The integrability of different $F(R)$ gravity models is actively investigated~\cite{Paliathanasis:2016tch,Paliathanasis:2011jq,Paliathanasis:2015aos,Papagiannopoulos:2018mez}. In particular, it has been shown that for some values of the constants $q$ and $n$ the cosmological equations of $R+qR^n$ model are integrable in the sense that they pass the singularity (Painleve) test~\cite{Paliathanasis:2016tch}.
We plan to generalize the proposed method on other modified gravity models, including the Palatine gravity and $F(R)$ models with an ideal gas~\cite{Papagiannopoulos:2018mez}.

We show that the behaviour of the Hubble parameter is essentially different in the Einstein and Jordan frames. In the Einstein frame, only monotonic behaviour is possible, whereas in the Jordan frame we have found both monotonic and nonmonotonic behaviour of the Hubble parameter. In particular, bounce solutions with a maximum of the Hubble parameter after bounce have been obtained in quadratures. Such behaviour may be suitable for inflationary scenarios. The considered $R^2$ model does not include the standard Einstein-Hilbert term and cannot be considered as a realistic model after inflation, but can be a good approximation at large values of $R$, when the $R^2$ term dominates.

{ \ }

\textbf{{Acknowledgements}}
\begin{acknowledgement}
We are delighted to thank A.A.~Starobinsky for his useful comments.
S.Yu.V. is partially supported by the Russian Foundation for Basic Research grant No.~20-02-00411.
\end{acknowledgement}

\end{document}